\documentclass[treatise,english,allpages]{ouvrage-hermes}[2005/11/14]

\usepackage{graphicx}

\title[%
Digital Materials]{%
Digital Materials}

\author{%
Samuel \Name{Forest}\\
Firstnameb \Name{Lastnameb}}

\date{%
April~5, 2016}

\makeindex

\begin{document}

\hbadness=2000
\emergencystretch=2em
\lefthyphenmin=3
\righthyphenmin=3

\frontmatter


\mainmatter

\PresentationOfAuthors{Chapter written by~}
\ChapterAuthor[
Dislocation Based Mechanics]{Dislocation Based Mechanics: the various contributions of Dislocation Dynamics simulations}{
Sylvain \Name{Queyreau}}

\section{Introduction}

Plastic deformation In crystalline materials is controlled by the motion and interactions of dislocations ~\cite{Anderson:2017}. Discrete Dislocation Dynamics (DDD) simulations have now existed for about 25 years to investigate plastic flow at the mesoscale, which lies at a critical position between two very different scales. At small scale, atomistic simulations are particularly adapted to investigate the core structure or mobility of individual dislocations, while continuous approaches can simulate the complete and continuous mechanical behaviour at the scale of the components. the purpose of DDD is to simulate the evolution of large dislocation ensembles in reaction to an external load and to assess the collective behaviour of dislocations. 

DDD accounts for both short- and long-ranged interactions between dislocations occurring within an evolving microstructure discretised in connected segments and embedded in a usually homogeneous matrix. DDD numerical efficiency resides in that stresses are not evaluated in the entire domain but rather only on the degrees of freedom of the dislocation microstructures. Basic properties of dislocations are provided from dislocation theory and/or atomistic data. In the past decade, a number of noticeable technical progresses have been accomplished by the DDD community such as, for example, the development of non-singular theories of dislocations ~\cite{Cai:2006, Po:2014}, accounting for anisotropic elasticity ~\cite{Aubry:2013, Aubry:2014,Lazar:2019}, tighter connections with atomistic simulations, and robust hybrid methods to solve finite domain plasticity problems ~\cite{Weygand:2002,Deng:2008,Vattre:2014,Jamond:2016,Bertin:2018}, among others. These technical progresses lead to significant advances in the understanding of dislocation-based plasticity of metals. 

This chapter aims at presenting a review of DDD technical developments and investigations from the past decade or so. We will mostly focuss also on 3D approaches, as 2D simulations are somewhat associated to a different physics. In the context of this book, special attention will also be provided to the link towards larger scales. Readers interested in the rich physics of dislocations are referred to ~\cite{Hull:2001, Kubin:2013, Anderson:2017}.

\section{Overview of Discrete Dislocation Dynamics}

The Dislocation Dynamics technique has been designed and improved over the years to investigate the collective behaviour and interactions of complex dislocation microstructures in reaction to an applied external loading. It appears therefore as a tool of choice to establish the link between microstructures and mechanical properties. The numerical competitiveness of DDD is twofold. First, since defect free regions of the materials play a little role in the plastic behaviour of -homogeneous- crystalline materials, modelling dislocations as an idealized network of connected lines contained in a continuous elastic medium has been recognised as an efficient strategy. Second, a number of analytical expressions and models exist which allows for a fast and precise evaluation of dislocation interactions\footnote{As we will see a bit further, analytical closed form expressions for energies and interactions of dislocations, in particular, exist for a handful of rather simple configurations, and other solution much be proposed for finite domain plasticity problems as discussed in section 1.2.4}. The derivation of analytical or semi-analytical expressions for dislocation interactions is indeed an active field of research in the DDD community (see section 1.2.3). An exhaustive list of DDD codes is provided in ~\cite{Kubin:2013} (at the end of Ch. 4).

In brief, the DDD technique simulates the evolution of discretized dislocations in reaction to the mechanical stresses and interactions that set them in motion. A general mobility function $M$ for a dislocations segment consists of the following form: 
\begin{eqnarray*}
\mathbf{v} = M\left(\mathbf{\sigma(x), b, \xi},T, material \right)
\end{eqnarray*}
and depends upon the dislocation type, character, temperature $T$ but mostly will be very material depend even if some general functional can be founds (see  section 1.2.2). The motion of dislocation also depends upon the local stress field that are applied $\sigma(\mathbf{x})$ onto them. While, glide is often constrained in a finite set of predefined crystallographic planes, the number of degrees of freedom (DOF) remains virtually infinite. As the numerical cost of force evaluations scales with the number of DOF, a compromise should be found between accurate description of the dislocation curvature and the number of discretised elements. Therefore, several discretization strategies have been proposed in the litterature. 
\begin{itemize} 
\item{In nodal based strategies \cite{Weygand:2002,Arsenlis:2007}, equations of motion are solved at $nodes$ as dislocations are described by straight $segments$ connected by $nodes$.  Forces felt by dislocations are integrated along the segments and transformed in a nodal force by means of shape functions. The gliding mobility of the node is averaged as the mobility of the two connected segments.   A drag coefficient accounting for the node deplacement along dislocation line -with no physical meaning- must be defined and must preserve the dislocation flexibility.}
\item{In $segment$ based codes \cite{Devincre:2011}, equations of motion are solved at the centre of straight segments, typically. To further reduce the number of possible DOF, segment orientations are also limited in number and are often chosen as special physical orientation such as, pure screw, edge, and junctions orientations. Non linear mobility function can be easily accounted for but the topology management is more complex than, for example, nodal approaches.}
\item{In $parametric$ approaches \cite{Ghoniem:2000, Po:2014}, dislocations are described by splines using a Galerkin approach. The use of non straight segment captures well the smooth flexibility of dislocations. The number of DOF required to describe a given dislocation curvature can potentially be smaller than other approaches. However, the development of analytical expressions is somewhat more cumbersome.}
\end{itemize}
Futhermore, the Burgers's law: $\Sigma_i \mathbf{b}_i = \mathbf{0}$ dictates that the sum of Burgers vectors $\mathbf{b}_i$ of segments entering a node should balance the  Burgers vectors of exiting segments. The microstructure should therefore consist of closed-loop geometries as dislocations ending freely in a crystal are unphysical and are associated with a stress divergence\footnote{On a theoretical standpoint, the stress divergence of an open loop configuration can however be removed by using rational stress expressions derived by \cite{Asano:1968} for linear elasticity.}. Ultimately, the choice of discretisation strategy is a matter of taste ~\cite{Cai:2006} and it should have a weak impact on the DDD results in comparison to critical choices regarding the physics included in the simulation.

\subsection{Initial configurations and Periodic Boundary Conditions}

As metastable defects, dislocations are -almost- always present in crystalline materials obtained by conventional processes. Initial dislocation microstructures are therefore typically introduced at the beginning of DDD simulations. Since the so-called Frank network adopted by dislocations initially is poorly known, initial dislocation microstructures are arbitrary. The original solution based on Frank-Read sources -associated to unphysical pinning points and divergent stress fields- is gradually abandoned in favour of closed geometries such as planar or prismatic loops. Prismatic loops consist of four successive -initially- edge segments, with two segments belonging to a primary system and two others to a deviated system. Interestingly, the length of the deviated segment has a huge impact on the initial strengthening, due to the very strong collinear annihilation existing between primary and deviated systems. When starting from prismatic loops of various geometries, the mechanical response nevertheless converges after some deformation as the density on the deviated system will naturally be regulated by cross-slip events and collinear annihilations. Some other authors prefer ~\cite{Aubry:2019} to use relatively small loops that will operate akin to sources. Depending on the strain rate, a marked yield point may be observed on the deformation curve, and the microstructure will quickly be very different from the initial one. 

In large scale DDD simulations, dislocation density increases naturally, without the recourse to artificial sources, by the temporary or permanent pinning of dislocations by obstacles. In real systems, the nucleation of fresh dislocations can occurs under special conditions from surface or under shock. Conventional DDD is not particularly suitable for the investigation of dislocation nucleation as it would require accounting for non-linear elasticity and precise description the dislocation cores. However, practical solutions have been proposed to address these issues. Ultimately, confidence in the arbitrary dislocation microstructures can be found in the  ability of DDD simulations to reproduce flow stresses, hardening and deformed dislocation patterns obtained experimentally on deformed single crystals (see section 1.3).

As the mean free path of dislocations can be much larger than the simulation dimensions, extended gliding planes are permitted by the use Periodic Boundary Conditions. The total dislocation density tensor over the whole simulation domain must be null, and PBC enforces a balanced flux of dislocations across the boundaries of the domain. To further lengthen the extended gliding plane offered to dislocations, the simulation domain can be rotated with respect to the crystal lattice. For segment based simulations, the size and orientation of the simulation basis must be carefully defined to respect the underlying simulation crystal lattice. Several approaches have been proposed to solve the associated Diophantine equations relating the simulation dimension and the projection of the PBC normals with respects to the lattice ~\cite{Madec:2004, Monnet:2006}. Interestingly, similar problems arise in atomistic simulations when two crystals are rotated with respect to the simulation box to form grain boundaries, have only recently been practically solved ~\cite{Banadaki:2015}.

\subsection{Mobility functions}

Mobility functions relate the effective shear strain felt by a given segment of dislocation to the resulting velocity ~\cite{Kubin:2013} . Mobility functions are crucial in describing the low temperature behaviour of materials, in particular. The effective stress  $\tau_{eff}$ is the total shear contribution felt by the dislocation in the gliding direction. Beyond the thermal regime and for conventional strain rates where relativistic effects are not felt, the velocity of dislocations $v$ simply reads: $|\mathbf{v}| = |\mathbf{b}| \tau_{eff} /B(T)$ where $B(T)$ is a dragging coefficient describing the phonon scattering induced by dislocation motion. In materials with large lattice friction e.g. bcc or hcp materials at low temperature, dislocation exhibit a much more complicated mobility. Taking bcc Fe as an example, athermal regime is known to be controlled by the slow motion of screw dislocation due to the large lattice friction. Screw dislocations move from one Peierls valley to the next, separated by $h_0$, through the so-called mechanism of double kink nucleation and propagation along the line if successful. The nucleation rate is generally the controlling mechanism. The resulting thermally activated behaviour is commonly modelled as an Arrhenius equation at a mesoscale as:
\begin{eqnarray}
|\mathbf{v}|  = h_0 {v_Db \over w} {L \over w} \exp \left( - {\Delta H(\tau*) \over k_B T}\right)
\end{eqnarray}
where $k_B$ is the Boltzmann constant, and the attempt frequency of a kink-pair of width w is $v_D |\mathbf{b}| /w$ with $v_D$ the Debye frequency.  ${L /w}$ where $L$ is the screw length represents the ratio of competing nucleation sites. The  
activation enthalpy $\Delta H$ replaces here the Gibbs activation energy as entropy is commonly neglected. $\Delta H$ is typically constant with temperature $\Delta H = C k_B T$. $\Delta H$ strongly decreases with the thermal stress $\tau*$, following the phenomenological expression ~\cite{Kocks:1975}:
\begin{eqnarray}
\Delta H = \Delta H_0 \left( 1 -\left( {\tau* \over \tau_P} \right)^p\right)^q
\end{eqnarray}
where $\tau_P$ is the Peierls stress and $p,q$ are exponents. Atomistic simulations are particularly well suited to defined mobility law e.g. ~\cite{Monnet:2004,Gilbert:2011}. Note also that dislocations with other that screw character may be thermally activated ~\cite{Monnet:2009,Queyreau:2011}. 

It must be emphasized that the dependance upon the dislocation length $L$ appearing the exponential prefactor, which has been been observed in situ within a TEM during deformation ~\cite{Louchet:1979, Caillard:2010, Caillard:2013}, is crucial in capturing the proper low temperature behaviour of bcc metals.

\subsection{Forces on dislocations}

Force evaluation on all segments of the dislocation microstructure is arguably the most demanding part of DDD simulations, despite the constant development of analytical and semi-analytical models. In this section, we will focus mostly on stress and force expressions that have been developed in the context of DDD simulations. Readers interested in a more complete list of analytical models for dislocations are refered to ~\cite{Anderson:2017}. Beyond the core region, dislocations generate only linear elastic distortions in the surrounding matrix. Therefore, the superposition principle applies. For the sake of simplicity, we start with the dislocation-dislocation interactions. When considering an infinite and homogeneous linear elastic medium, a first dislocation loop generates a stress field  $\mathbf{\sigma}^{\infty}(\mathbf{x})$ at a
point $\mathbf{x}$ defined as a contour integral along the loop ~\cite{Mura:1987}:
\begin{eqnarray}
\sigma_{ij}^{\infty}(\mathbf{x}) &=& C_{ijkl} \oint \epsilon_{lnh} \, 
C_{pqmn} \, { G_{kp,q} (\mathbf{x-x'}) }\, b'_m \, dx_h' \label{eq:sijmura} \\
G_{ij} \mathbf{(x-x')}&=&{1 \over 8\pi\mu} \left[ \delta_{ij} \partial_p \partial_p  R - {1 \over 2(1-\nu)} \partial_i \partial_j R \right]
\end{eqnarray}
Where $C_{ijkl}$ is the elastic stiffness matrix of the medium,  $\epsilon_{lnh}$ is the permutation operator. $\mathbf{b'}$ is the Burgers vector of the loop and $\mathbf{x'}$ is the
coordinate that runs along the dislocation. $R$ is the metric: $||\mathbf{x-x'}||$. $G_{kp} (\mathbf{x-x'})$ are the fundamental Green's functions of elasticity, here given for isotropic elasticity. 
They correspond to the displacement component in the $x_k$ direction at
point $\mathbf{x}$ induced by a body force located at $\mathbf{x'}$ in the $x_p$
direction ~\cite{Mura:1987}. If $\mathbf{x}$ now spans a second dislocation, this gives rise to
the so-called Peach and Koehler force:
\begin{eqnarray}
\mathbf{F}^{PK} = \oint N(\mathbf{x}) (\sigma_{ij}^{\infty}(\mathbf{x}) \cdot  \mathbf{b})\times \mathbf{\xi} \, dL
\end{eqnarray}
Where $N(\mathbf{x})$ are potential shape functions to distribute the force on nodes, and $dL$ is a length along the second dislocation.  $\mathbf{b}$ and  $\mathbf{\xi}$ are the Burgers and line vectors of the second loop, respectively. Closed form analytical expressions for the force integrals can found in ~\cite{Devincre:1995, Arsenlis:2007} for straight segments and spline segments in ~\cite{Po:2014}.

The conventional theory of dislocations is associated to a well-known singularity of the stress field close to the dislocation core. This is imputed to the sharp (Dirac) distribution of Burgers vectors used in to calculate the elastic fields. While other regularisations have been proposed to remove the singularity ~\cite{Anderson:2017,Brown:1964, Gavazza:1976}, the solution provided by Cai {{\it et al.}}~\cite{Cai:2006, Arsenlis:2007} leads to fully-analytical and compact force expressions for the double integral above and consisted in replacing the sharp Burgers distribution by an isotropic spreading function and assuming linear elasticity still applied at the dislocation core. The metric $R$ is replaced by $R_a = R*w(\mathbf{x})$ its convolution with the core distribution $\mathbf{b}w(\mathbf{x})$. An appropriate choice of distribution lead to a very simple equation for $R_a = \sqrt{\mathbf{R \cdot R} + a^2} $  and few additional higher order terms appearing in the stress expressions when compared to the singular theory. As the model was derived for dislocation-dislocation interaction, $w(\mathbf{x})$ is in fact a double convolution of the Burgers distribution, and only an approximate distribution is provided for the single convoluted distribution. This could potentially lead to smoother fields close to dislocation core and double convolution is, in principle, not required when calculating stress at a point in the matrix. This method has only an approximate description of the core.

It must be noted that nodal force expressions identical to the ones derived from equation [1.5] can be obtained by a different route starting from the mechanical work of interaction existing between the two segments ~\cite{Mura:1987}: 
\begin{eqnarray}
E_{el} &=& {\mu \over 16 \pi} \oint \oint b_i b'_j \partial_k \partial_k R_a dx_i dx'_j - {\mu \over 8 \pi} \oint \oint \epsilon_{ijq} \epsilon_{mnq} b_i b'_j \partial_k \partial_k R_a dx_m dx'_n \nonumber \\
&& + {\mu \over {8\pi (1-\nu)} } \oint \oint \epsilon_{ikl} \epsilon_{jmn} b_k b'_m \partial_i \partial_j R_a dx_l dx'_n
\end{eqnarray}
A conservative force corresponding to self or mutual interactions can be then defined at dislocation node $\mathbf{q}$ as:
\begin{eqnarray}
\mathbf{F}_i &=& - {\partial  E_{el} \over \partial \mathbf{q}_i}
\end{eqnarray} 

Another contribution is the so-called self interaction of the segment ~\cite{Brown:1964, Gavazza:1976}. As for distinct dislocation interactions, the self force can be defined in a self-consistent fashion as the negative of gradient of self energy with respect to, say one of the segment node. The self energy corresponds to the double line integral of ver the loop or equally the volume integral of the free energy in the volume and is therefore considered as non local.  In Cai et al. theory ~\cite{Cai:2006}, when taking the gradient of the self energy, a part of the energy is coming from contribution within the spread radius $a$. This is a core contribution of the model, but it is not $per\, say$ the core energy of a dislocation as there is no information from atomistics (see e.g. ~\cite{Clouet:2011} ) and linear elasticity is used. Since, line tension constitutes ~\cite{ DeWit:1959} the leading term in the self force, line tension expressions or the heuristic based approach from Brown ~\cite{Brown:1964} are commonly employed in DDD simulations. 

The case of anisotropic elasticity is typically more difficult as closed form analytical solutions for the stress field around dislocations do not exist. Elastic anisotropy affects remote dislocation interactions and may lead to strong changes in equilibrium shape of loops ~\cite{Bacon:1979, Aubry:2011}. The derivative of the required Green's functions can however be expressed as angular integrals around dislocation segments ~\cite{Mura:1987}. Aubry and Arsenlis ~\cite{Aubry:2013} approximated these integrals by means of spherical harmonics expansion, the order of which is truncated and adapted to the targeted precision. Truncation order must be increased as anisotropy coefficient increases. Nonetheless, this semi-analytical approach is a significant improvement over the brute force double integration to obtain interacting forces, and allowed a reduction of the extra cost of full anisotropy from more than two orders of magnitude to somewhere between same order to an order more expansive that the classical isotropic solution ~\cite{Aubry:2013}. Hybrid DD-FFT solution is a promising method and is suggested as a strong competitor to fully anisotropic calculations, however the stress field is obtained on a grid in volume and the same control of numerical errors has still to seen. When not looking for the full anisotropic machinery, effective isotropic elastic constants and tabulated line tension can provide estimates of the behaviour of curved dislocations in anisotropic medium ~\cite{Bacon:1979, Aubry:2011, Kubin:2013, Madec:2017}. 

Lazar and coworkers \cite{Lazar:2012,Lazar:2013,Po:2014} have formulated a non-singular theory for dislocations based on strain gradient elasticity of Helmholtz type. A characteristic internal length $l$ is introduced and associated to higher stress terms. The displacement and stress fields are non-singular thanks to the use of an isotropic spreading function $A(R)$ of the Burgers vector simply defined as the convolution of $A(R) = R*G$, where $G$ is now the isotropic Green's function of the Helmholtz operator. It yields $A(R) = R+{2l^2\over R}(1-\exp^{-R/l})$. Expectedly, the internal length $l$ was found to be slightly below $1|\mathbf{b}|$ from atomistics in W, which is close to the ideal isotropic material.  In contrast to \cite{Cai:2006}, the Cauchy stress is defined from a single convolution with the core spreading function. Very recently, Lazar and coworkers ~\cite{Lazar:2019} extended their work to the cases of nonlocal anisotropic elasticity of generalized Helmholtz type. The theory is now associated to six internal lengths and  the dislocation core distribution is now anisotropic. The line integrals along dislocations loops and angular integrals appearing in the Green tensor need however to be evaluated by numerical quadrature.

Naturally, other interactions with dislocations can be included starting from equation 1.5 and a stress field expression or from equation 1.7 and an interaction energy,  and the superposition principle, whether preference is given to stress or interaction energy. This is the case for the external applied stress, or the interactions with Eshelby inclusions that can be introduced to model second phase particles ~\cite{Aubry:2019} or punctual defects ~\cite{Cai:2014, Gu:2018}. 

\subsection{Topological changes}

During the simulation, the dislocation microstructure strongly evolves and a number a topological changes need to be addressed.
\begin{itemize}
\item{Dislocation lines may be rediscretized to adapt the segment length to the local curvature. Different criterion has been proposed to this aim ~\cite{Devincre:2011,Arsenlis:2007}}.
\item{Different part of the dislocation microstructure may collide during the motion of mobile segments. If the Burgers vectors are different this may lead to the formation of a reaction, that typically minimizes the elastic strain energy and the resulting dislocation may well be sessile in any of the initial planes. The formation and destruction of the reactions occur through zipping and unzipping by the lateral movement of the mobile segments. This plays an important role in the deformation of crystalline materials as reactions are almost always operative and they even constitute the most part of hardening in pure materials in the athermal regime.}
\item{Finally, screw dislocations have the possibility to change of the possible slip planes in zones that are offered to them. In fcc crystals, \emph{Cross Slip} (CS) occurs through the so-called Escaig-Friedel mechanism, according to which dissociated partial dislocations spread in a primary plane must recombine prior to any change of plane. This core transformation is necessarily thermally activated and with a power law dependance of the activation energy with respect to the stress and a length dependence of the prefactor. In DDD, the probability for a CS event to occur is handled by a Kinetic Monte Carlo algorithm. In bcc Iron, dislocation cores are compact ~\cite{Frederiksen:2003, Monnet:2004} and multiple slip planes are possible, CS is commonly considered as very easy in agreement with the wavy slip lines observed at sample surface. A purely mechanical criterion was used in ~\cite{Queyreau:2008}. }
\end{itemize}

\subsection{Boundary conditions}

\begin{figure}[h]
\includegraphics[scale=0.5,bb=0 0 100 100]{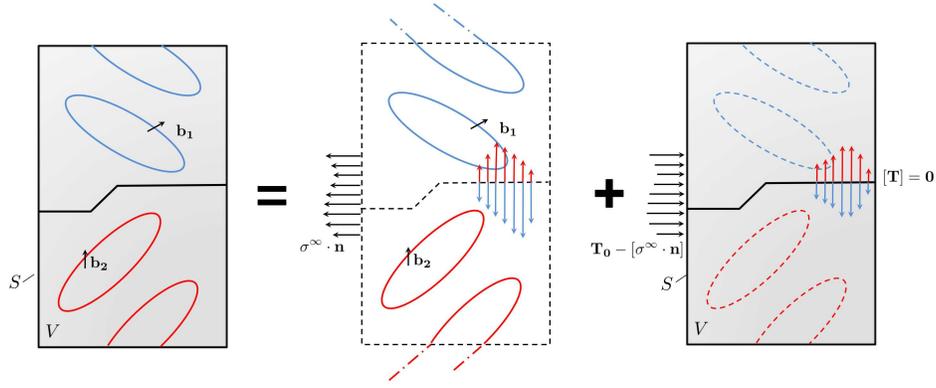}
\caption{ Illustration of hybrid method solve
  from~\cite{Queyreau:2014} (a). Two slip families of Burgers vectors
  $\mathbf{b_1}$ and $\mathbf{b_2}$ are present in a {bicrystal}  $V$, bounded by $S$. (b) The traction field due to the
  discrete dislocation ensemble in the infinite domain is evaluated on
   the domain internal and external {boundaries}. This calculation is performed on
  dislocations contained within the domain and on virtual segments used
  to close open dislocation loops. (c) Then, a FE or BE simulation of
  a non-dislocated elastic medium gives the corrective stress field
  that enforces the desired boundary conditions, i.e. set traction values $\mathbf{T_0}$ at
   external surfaces and no traction discontinuity at internal interfaces.}
\label{geom}
\end{figure}

The numerical efficiency of DDD simulations lies in its efficient and accurate evaluation of dislocation interactions. This is permitted by the use of analytical expressions for the dislocation-dislocation interactions, for example, that have been formulated in the case of few simple configurations, typically the case of closed loops contained in an infinite and homogenous domain. The use of PBC conditions mentioned earlier, perfectly matches this situation. However, this is not the case when considering finite domain geometries or internal interfaces or inhomogeneities. To address this, hybrid solution associating DDD with FEM or BEM have been proposed to correct -or partially replace- DDD fields ~\cite{Giessen:1995, Lemarchand:2001, Martinez:2002, Zbib:2001, ODay:2004, Weygand:2002,Deng:2008 Tang:2006, Weinberger:2009, Vattre:2014,
  Jamond:2016, Bertin:2018}. The original solution was proposed by Needlemann and Van der Giessen and is based on superposition principle ~\cite{Giessen:1995}
The stress field solution is decomposed into a "classical"
 ($\infty$)  field associated to dislocations in an infinite domain and a
corrective  ($\,\hat{}\,$)   field that enforces the desired boundary
conditions at the internal or external boundaries of the simulated
domain (see figure 1.1).
\begin{eqnarray}
\mathbf{u = u^{\infty}+\hat{u}} \\
\mathbf{\sigma = \sigma^{\infty}+\hat{\sigma}}
\end{eqnarray}
The ($\infty$) infinte domain fields do not respect the desired boundary conditions, the corrective field are obtained from a FE or BE simulations carried over a non-dislocated medium with the following boundary conditions:
\begin{eqnarray}
\mathbf{\hat{\sigma} \cdot n} =\mathbf{\hat{T} = T_0-T^{\infty}} \mbox{\, on\, $S_T$}\\
\mathbf{\hat{u}}=\mathbf{ u_0-u^{\infty}} \mbox{\, on\, $S_u$}
\end{eqnarray}

An alternative hybrid solution called Discrete Continuum Model (DCM) was proposed in a series of papers ~\cite{Lemarchand:2001,Vattre:2014,Jamond:2016}. The coupling is also rather strong, with the two codes handling the same domain and similar timesteps. DDD manages short range interactions between dislocations contained in an element, while FEM handles long range dislocation interactions among dislocation in different elements and boundary conditions. The long range interaction is performed using an eingenstrain formalism ~\cite{Mura:1987} and a regularization procedure of the plastic deformation induced by gliding area swept by dislocation in DDD and the corresponding plastic strain on the integration points of the FEM mesh. This method can be seen as a FEM calculation where the constitutive equations of crystalline plasticity are replaced by a synchronous DDD calculation. 

The evaluation  of the unbalanced tractions induced by
dislocations at the boundaries of the
domain~\cite{Weygand:2002,Deng:2008} is required prior to the FE/BE. The
tractions $\mathbf{\sigma}^{\infty} \cdot \mathbf{n}$ are expressed as the triple
integration of the infinite domain stress field $\mathbf{\sigma}^{\infty}$ along
the dislocation line and over the surface element of normal $\mathbf{
  n}$.
Tractions are then projected and appropriately distributed as nodal
forces $\mathbf{F}^{(n)} $ over the nodes of the surface element $S^{(E)}$ by means
of shape functions $N$:
\begin{eqnarray}
\mathbf{F}^{(n)} &=& {\int_{S^{(E)}} [\mathbf{\sigma^{\infty}(x) \cdot n}] \, N^{(n)}(\mathbf{x}) \, dS}. 
\end{eqnarray}

The evaluation of unbalanced tractions induced by DDD stress fields is arguably the most demanding task when associating DDD to
FE/BE as this operation must be carried out over all segments constituting the microstructure and on all surface elements delimiting
the domain~\cite{Weygand:2002,El-Awady:2008}.  In ~\cite{Queyreau:2014, Queyreau:2019} 
Queyreau and coworkers proposed an analytical solution in the case of the non-singular theory of 
Cai and coworkers ~\cite{Cai:2006}. These analytical solutions are obtained by sequences of integration by parts that exhibit recurrence relations, and are found to be very precise (as analytical) and much more efficient numerically that classical double numerical integration strategies. This model was then extended in ~\cite{Liu:2016} to the case of anisotropic elasticity ~\cite{Aubry:2013}. 

Enforcing the proper boundary conditions at interfaces lead to significative modifications to the infinite domain solutions, leading to image stresses felt by crystal dislocations. This particular boundary value problem in presence of dislocation can only be assessed by hybrid DDD-FEM solver.  These corrections are particularly important in the case of anisotropic elasticity, which explains why conventional DDD is still commonly employed in the case of polycrystals in the case of isotropic elastic medium. However, plastic deformation is intrinsically anisotropic and different strains can results from different dislocation activity in adjoining crystals, which are rotated with respect to each other. As matter must remain compatible across the interface, this plastic incompatibility at the interface may be accommodated by elastic deformations in both grains ~\cite{Gemperlova:1989}. Contrary to the elastic incompatibiltiies induced by different microstructures on both sides of the GB, plastic incompatibility are uniform and lead to long range incompatibility stress fields ~\cite{Sutton:1995}. $Sigma \tau_i$ Finally, a third ranged contribution may be found in the stress induced by the GB defect structure itself ~\cite{Sutton:1995}. From the seminal the works of Read and Shockley and Frank ~\cite{Sutton:1995}, low angle grain boundaries are commonly described as arrays of dislocations. These arrays are particularly simple for symmetric pure tilt and twist grain boundaries. In the case of semi-coherent interfaces, misfit dislocations may accommodate the large 
 lattice and elastic mismatch. As the spacing between GB dislocation decreases with misorientation angle according to Frank's equations, for HAGB, GB dislocations core overlap and cannot be individualised. This is also the case for mixed tilt twist GB. Investigating the GB structure has motivated a wealth of atomistic study. It is however striking, that the atomistic energy of so many GB can be fitted by a RS type equation in the 5D parametric space of GB ~\cite{Bulatov:2014,Beers:2015}.

Formally, it is possible to determine possible dislocation arrays from the transformation related the two adjoining crystals.  For planar interfaces, 
 possible arrays of misfit dislocations are 
 expected to comply with the so-called quantized Frank-Bilby equation: 
 $\mathbf{B = (F_M^{-1} - F_P^{-1} )\cdot k}$ that relates the displacement gradient $\mathbf{F_M}$ and $\mathbf{F_M}$
 for the matrix M and particle P with respect to a reference interface state, to the sum 
 of interface Burgers vectors  $\mathbf{B}$ crossed by an interface probing vector $\mathbf{k}$.
  In theory, the number of possible arrays of dislocations
 capable of accommodating  is not unique, and requires to find the reference state of 
 the interface. Vattr\'e and Demkovitz ~[\cite{Vattre:2013,Vattre:2015,Vattre:2016}] devised a strategy in a number of papers to define 
 a unique solution for semi-coherent interfaces
 in a bycristal, by testing difference reference states and using continuous mechanics and the Stroh 
 anisotropic elasticity formulation. The dislocation array  solution is then identified as the one 
 cancelling far field stress field induced by the interface and associated to the lowest energy.

\section{Mesoscale plasticity}

In this section, we review a number of recent contributions from the DDD community, to the understanding of the plastic behaviour of crystalline materials. The sections focusses primarilly on pure metallic systems, of mostly cubic lattice structure. Other recent reviews can be found elsewhere regarding strengthening in alloys ~\cite{Aubry:2019} or irradiated materials  ~\cite{Marian:2018, Monnet:2018}.

\subsection{Forest interaction}

The so-called Forest interactions occurring between dislocations plays a central role in the plastic flow of crystalline materials as dislocation-dislocation interactions are almost alway operating and it helped shaping up strengthening models for other dislocation-obstacle interactions. Assuming that the Schmid law is respected, the critical stress $\tau_c$ required to activate a slip system 'i' is given as ~\cite{Franciosi:1980, Kocks:2003}: 
\begin{eqnarray}
\tau_c^i = \mu b  \sqrt{\Sigma_j a_{ij} \rho_j}
\end{eqnarray}
where $\rho_j$ is dislocation density attributed to system 'j' and $a_{ij}$ is the so-called interaction matrix that gathers the coefficients measuring the strength of the interaction between 'i' and 'j'. $a_{ij}$ are poorly known from experiments ~\cite{Franciosi:1980} and the determination of coefficients $a_{ij}$ of the interaction matrix from DDD is not an easy task $a\, priori$.  Indeed, the length, density, character of the interacting dislocations and elastic properties can affect the reaction stability and these parameters should be averaged in a statistical manner. DDD simulations modeling the latent hardening test is now a common approach to achieve this, and consist of an active slip system that interacts with one or several inactive slip systems leading to a single reaction. While having an inactive slip system may in principle lead to different dislocation patterns and therefore different sampling of dislocation characters and lengths, the coefficients determined this way are self consistent with the DD simulations where all reactions are included and robust enough for the prediction of the plastic anisotropy of single crystals (see below). This approach has been employed for  fcc ~\cite{Devincre:2006, Madec:2008, Madec:2017}, bcc ~\cite{Madec:2004, Madec:2008,Queyreau:2009,Madec:2017} and hcp materials ~\cite{Devincre:2013,Bertin:2014}. 

As the largest part of the forest strengthening is related to the short range interactions, as opposed to remote interactions, equation 1.9 finds its physical origin in the line tension to bow out the mobile segment pinned by a pair or reactions ~\cite{Saada:1960}. Therefore, the simple expression above neglects the logarithmic dependence of the line tension and $a_{ij}$ thus depend upon the dislocation density \cite{(Basinski:1979,Madec:2002,Devincre:2006}. Devincre and coworkers ~\cite{Devincre:2006, Kubin:2008} proposed to correct $a_{ij}$ from this effect as:  $ a_{ij}(\rho_j)  = a_{ij}^0 \left(\ln(1/b\sqrt{(a_{ij}^0 \rho_j )} ) \big/  \ln(1/b\sqrt{(a_{ij}^0 \rho_j^0 )} )  \right)^2$, where $\rho_j^0$ is the reference density at which coefficient $a_{ij}^0$ was determined, and $1/b\sqrt{(a_{ij}^0 \rho_j )}$ is an approximation of the average dislocation spacing. This correction was latter refined as the logarithmic correction should apply to the part of interaction strength related to contact reaction ($\approx 80 \%$) and another constant contribution comes from long range interactions ($\approx 20 \%$) ~\cite{Devincre:2006,Kubin:2008}. 

Junctions - and their corresponding interaction coefficients- are typically weaker for bcc metals ~\cite{Madec:2004,Madec:2008,Queyreau:2009,Madec:2017} than for fcc ~\cite{Devincre:2006, Madec:2008, Madec:2017}.  Among reactions, the collinear reaction that may occur between two distinct slip systems with same Burgers vector is of particular interest. First, while this reaction leads to the annihilation of the interacting segments, the effective strengthening obtained is by far the strongest among reactions as primary segments may move far from the collision point and many loops and debris are left behind. Contrary to other contact reactions, collinear annihilation does not lead to $direct$ dislocation storage and can lead to unstable activity on active slip systems. This could explain the slip system selection existing in hight symmetry loading axis where non collinear slip systems are activated. Second, collinear annihilation goes hand in hand with cross slip activity as screw dislocation change of plane for a collinear system. To the first order, reaction stability scales with the gain of elastic strain energy associated with the initially separated segments or with configuration where the reaction is formed. Neglecting the character dependance and the logarithmic cut offs, elastic strain energy of one segment roughly scales with $\mu |\mathbf{b}|^2$.

CS appears more ambivalent that the recovery mechanism it is commonly considered to be. At small to intermediate strain, CS along the creation of collinear annihilation, was instrumental in capturing the correct $stage \, I$ hardening of single crystals with an orientation such that a single slip system is activated ~\cite{Devincre:2007}. The apparition of typical microstructures made of entanglements is also intimately related to collinear annihilations. In bcc Fe, where CS can be considered to be very easy, the hardening effect of CS is even more apparent ~\cite{Queyreau:2008} with the production of many debris and and a large hardening rate in stage I. However, at larger stresses CS becomes an effective way to by-pass strong obstacles or stress concentration, as soon as the the external stress is sufficient to activate large segments in the deviated system, and the recovery effect of CS is recovered ~\cite{Kubin:2009}. This will depend upon stress, obstacle density and loading directions as it affects the Schmid factor on the deviated system. 

With an average length of mobile segment scaling as $k / \sqrt{a_{ij} \rho_j}$ ( $\approx$75\%, see below), binary junctions certainly represent the main portion of reactions occurring at least in the beginning of the deformation. More complex configurations have nonetheless drawn attention recently. For example multijunctions, may occur between a mobile segments and an already existing binary junction. As the reaction stability scales with the amount of elastic strain energy that is gained by the junction formation, these ternary objects may be very stable and strong pinning points. In ~\cite{Bulatov:2006}, Bulatov and coworkers provided that multijunctions could be observed in atomistic simulations and some TEM observations. These ternary objects can also be found in spades in large scale DDD simulations in bcc Mo, they may act as strong anchoring points to the dislocation microstructures and operate as FR-like permanent sources leading to a noticeable extra hardening to binary junctions, at least for latter deformation stages ~\cite{Bulatov:2006}. 

Madec and Kubin ~\cite{Madec:2008} investigated in details crystallography of ternary reactions. In the case of bcc lattice, the multiplicity of \{110\} and \{112\} planes leads to an incommensurable number of ternary reactions. Instead, the authors focussed their attention onto the set of 16 reactions -with various geometries and characteres- that can be found when loading single crystals along high symmetry axises. Ten of these are axial reactions, in line with the parent binary junction, and six others are called zigzag reactions, which exhibit a different line orientation than the parent binary junction. Using the same equation as for binary junctions, strength of ternary reactions was assessed from model latent hardening simulations including an additional active slip system. Ternary reactions -especially delocalised zigzag configurations- can be stronger than binary junctions. In contrast, with a lower number of slip planes in fcc and putting aside coplanar reactions, three ternary zigzag reactions were considered and involve a glissile segment. As ternary reactions in fcc are very similar to the structure of binary reactions, they also exhibit slightly the same strength. Formally, one could imagine rewriting the critical stress ~\cite{Franciosi:1982} as  $\tau_c^i = \mu b  \sqrt{\Sigma_j a_{ij} \rho_j + \Sigma_j  \Sigma_l  h_{ijl} \rho_j \rho_l} $ to capture higher order interactions. $h_{ijl}$ are now ternary interaction coefficients between system $i,j$ and $k$. 

When considering isotropic elasticity, interaction coefficients depend solely on the crystal structure and nature of the reaction formed. This is however not the case when considering anisotropic elasticity. Madec and Kubin ~\cite{Madec:2017} recently probed this effect in the case of binary reactions in several fcc (Ge, Al, Ni, Cu, Ag, Au) and bcc (Mo, Nb, W, Ta, Fe) materials at the onset of the athermal regime. Effective isotropic elastic properties are estimated from ~\cite{Bacon:1979}. For fcc materials, glissile junctions are now associated to a geometrically asymmetric configuration regarding the mobile segments orientation and the junction mechanical stability. $a_{ij}$ and $a_{ji}$ now differ when active and forest slip system are swapped. $a_{ij}$ Coefficient differ slightly from previous calculation (give relative variations) and scale linearly with the Poisson's ratio. The picture is somewhat much more complicated for bcc materials with no such simple evolutions of the coefficients. More reactions existing between {110} and {112} slip systems are now asymmetrical. 

Glissile reaction have again recently attracted some attention. Striker and Weygand ~\cite{Stricker:2015} recently assessed the impact of glissile junctions in the plastic behaviour of micropilars. Similarly to CS segments or pairs of multijunctions, glissile reactions can also operate as FR-like sources if the local stress is sufficient. the authors developed tools to track the plastic activity imputed to glissile junctions over a wide range of initial microstructure realisations and simulation sizes $D$ up to 2 $\mu$m. The contribution of glissile segments to the overall plastic deformation was found to be in the range  10\% to 50\% and roughly scaling with $D\sqrt{\rho}$. However, the glissile contribution is certainly much smaller in large crystals. Activation of a segment of length $L$ akin to a FR source scales with $1/L$. Binary junctions typically exhibit an average lenth $1/ \sqrt{a_{ij} \rho_j}$ which is much smaller than mobile $parent$ segments ~\cite{Devincre:2006}. Besides, the presence of free surface is strongly attractive to dislocation in the bulk leading to a strong annihilation and possibly to dislocation source starvation.

\subsection{Statistical investigations of dislocation mechanisms}

Most large-scale DDD simulations - including those discussed in this chapter - aim to sample the wide variety of dislocation interactions and elementary mechanisms to provide an averaged and representative behaviour of the simulated material. In this and the following section, we highlight recent studies that evaluate the variety of dislocation configurations and corresponding distributions. A first example of this type of work was conducted by Devincre and coworkers ~\cite{Devincre:2006} who determined distributions of junctions lengths $l_j$ and mobile segment lengths on the primary system $l_p$. This was achieved on model simulations of latent hardening (see previous section) and [111] and [001] multislip simulations of fcc Cu with various Forrest dislocation density $\rho_f$. In model simulations, junction length increases with the length of forest segments ~\cite{Queyreau:2009} until reaching a plateau of the order $<l_j> \approx 0.25 \rho_f^{-1/2}$ ~\cite{Devincre:2006} that is independent of the nature of junctions. The average mobile segment length however was found to scale with the strength of the junction as $<l_p> \approx 1/\sqrt{a_f \rho_f}$ where $a_f$ is the interaction coefficient.  Recently, Sills and coworker ~\cite{Sills:2018} proposed a similar analysis for a [001] multislip simulation in Cu. These authors observed a distribution of primary segments that is mostly consistent with an exponential distribution of the form: $p(l_p) = (1/<l_p>) \exp\left({- l_p / <l_p> }\right)$. With a an averaged $<l_p>$ proportional to $1/\sqrt{\rho}$. The authors explain the exponential distribution of $l_p$ by the fact that the probability to obtain a collision scales with a given mobile segment length $l$. Neglecting the junction length, the two new mobile segments created around the junction splits the length into $l_1$ and $l-l_1$. The junction can then be seen as a 1D Poisson process. The authors define two dimentionless coefficients: $\phi = N^2 / \rho$ with $N$ the number of segments, and $\beta$ the rate of stable junction formation, and recovered the classical rate equation for dislocation storage ~\cite{Kocks:2003}.

Another ingredient entering in hardening of materials with deformation lies thus in the storage rate of dislocations $d\rho$ with deformation $d\gamma$. Following the seminal work of Kocks and Meckin ~\cite{Kocks:2003}, storage rate on system 'i' in pure fcc materials in the athermal regimes is :
\begin{eqnarray}
 {d\rho_i \over d\gamma_i } = {1\over b} \left( {{1 \over L} -y_c \rho_i } \right)
\end{eqnarray}
where $L$ is the mean free path of mobile dislocation before being temporary or permanently locked in the microstructure and $y_c$ is the annihilation distance for screws controlling in part the dynamic recovery of $stage\, III$ (see ~\cite{Kocks:2003, Kubin:2009} for more details). The MFP of dislocations is typically poorly known from experiments and is commonly expressed as $L  = \sqrt{\Sigma a_{ij} \rho_j} / K_{hkl}$. This physical quantity depends on the number of active slip systems and thus the loading direction. In ~\cite{Devincre:2008}, the MFP is broken into three different physically-based parameters that are defined from averages on large scale simulations. These parameters corresponds namely to:  i) the rate of formation of stable junctions close to $p_o$ close to 10\%, and the two average lengths in the shape of segments  $<l> = k_0 /\sqrt{\bar{a} \rho}$  and the junction segment $<l_j> = \kappa <l>$ that are stored in the process, with $k_0 = 1.08$ and $\kappa = 0.29$. Interestingly, Sills and coworkers arrived to the same storage functional as Kocks and Meckin throught a different route. For this, the author also define a rate of formation of stable junction and a pinning point spacing as $\lambda = 1/\sqrt{0.55 \rho} $. The storage rate is then defined from exponential distribution of segment length described above. The full CP model proposed by Devincre and coworkers ~\cite{Devincre:2008} and parametrized solely from DDD was capable of reproducing most of the features of single-crystals deformed monotonously. This work thus demonstrate the DDD predictive capability and weak influence of the dislocation patterns, to the first order.

In a series of paper El-Azab, Deng and coworkers performed an extensive statistical analysis of time and spatial correlations of dislocation ensembles ~\cite{Deng:2007, Deng:2010, Xia:2016} with the ambition in mind to provide a continuum dislocation dynamics descriptions based on DDD results. Similarly to Crystal Plasticity, CDD aims at providing a continuous description of a dislocation system and its evolution by means of dislocation densities, but contains additional physical ingredients e.g. lines orientation or line curvature to capture more naturally the increase in line length and possibly reproduce dislocation patterns. To this aim, time and spatial corse graining must be carefully performed. Interested readers are referred to other chapters on continuous dislocation fields theories in the present book. in ~\cite{Deng:2007}, Deng and El-Azab determined from DDD the pair correlations existing during the deformation of Mo single-crystals deformed along  [001]. The radial correlation existing between pairs of segments of the same slip systems is peaked at $\mathbf{r_{ij}}| -> 0$ due to the line connectivity and displays oscillation indicating preferred spacing distances. The repartition in theses preferred spacings decreases with the separation distance and these amplitude oscillations in the correlation functions decrease with strain. The pair correlation is also strongly anisotropic as it depends upon the two angles necessary to orient the dislocation line. Dislocations are correlated at short distance, again due to connectivity and can be correlated or anti-correlated at medium distance, and uncorrelated or anti-correlated at long range. The authors suggest that the behaviour at intermediate can be due to the distance interactions between dislocations that may be attractive or repulsive. Spatial correlations among  pairs of segments belonging to different slip systems exhibit similar features, with a peak at close range and oscillations at preferred distances that decrease in intensity with strain. The correlation patterns are found to be very anisotropic. Rationalising pair correlations remain however a difficult task due to the large number of slip systems and geometric and physical parameters affecting correlations ~\cite{Deng:2010}. 

in CDD, density evolution is controlled by the rate of reactions $R$ and cross-slip events, that are seen as stochastic processes. Defining a relevant mesoscopic time scale at which these processes can be seen as continuous becomes key.  Deng and El-Azab ~\cite{Deng:2010},  investigated the temporal statistics and the rates of reactions (junctions and annihilations) and cross-slip events. For this, the authors used $marked\, point\, process$ and $time\, series$ to model the statistics observed in the DDD data. The correlation times for junctions and annihilation are found to be of the same magnitude and much smaller that correlation time for CS events. The rate of junction events was found to exhibit strong fluctuations that are superposed to a low background activity slowly increasing with strain. The large fluctuations are particularly  present at small strain. As two successive events within a correlation time are seen as correlated, shorter correlation time means events are independent. Interestingly, the probability distributions $p(R)$ of rate parameter also follows a power law scaling as $\approx R^{-k}$ with $k\approx 2 \pm 0.5$ (see a bit below). The analysis of CS events was then refined in ~\cite{Xia:2016} where CS rates are not considered as stationery but rather increase with strain. Indeed, the local stress fluctuations building up with accumulated strain, will eventually favour CS events \cite{Queyreau:2008, Xia:2016}. The coarse grained rate for CS events now includes an additional stockastic component to reproduce fluctuations, and is accounted for in CDD by a Monte Carlo scheme. This solution ultimately made possible to reproduce the $\mu b / \tau_c$ scaling of the deformation cells (while failing in other quantitative areas).

\subsection{Lattice friction and combination with other strengthening mechanisms}

A number of DDD investigations of the thermal regime of materials with large lattice friction exist in the literature. While many of them include a thermally activated character of screw mobility corresponding to DK nucleation and propagation, only few of them account for the dislocation length dependence that is instrumental in capturing the thermal regime of plastic flow ~\cite{Tang:1998, Monnet:2004, Naamane:2010, Monnet:2011}. In ~\cite{Naamane:2010}, such mobility law was parametrized from experiments conducted on single crystals of pure $\alpha$-Fe or $\alpha$-Fe with small amount of impurities, and focussing on \{110\} systems. Naamane and collaborators provided great care in the definition if the onset of plastic flow, which can be very different from the engineering estimate owing to the presence of an extended $stage\, 0$. During this initial stage, edge dislocations that are very mobile, produce most of the plastic deformation leading to an imperceptible hardening until they annihilate, become blocked, or exit the crystal. At the end of this stage, the microstructure is almost essentially composed of screw dislocations with a slow and thermally activated mobility, fitting the conventional picture of low temperature plasticity ~\cite{Louchet:1979, Caillard:2013, Kubin:2013}. At the mesoscale of DDD simulations, the backward jumps of DK cannot be neglected and the exponential form of equation [1.1] may include a hyperbolic sine function. For reasons of numerical efficiency, mobility functions of non-screw dislocations in ~\cite{Naamane:2010} is set to be proportional to the mobility of screw dislocation (and is independent of non-screw length). The ratio between non-screw and screw characters is set such as $k\,  >>\, 1$, with $k$ increasing as the temperature decreases. These ingredients allowed to reproduce the onset of plastic flow of Fe single crystals and the evolution dislocation microstructures with decreasing temperatures.

Monnet and coworkers ~\cite{Monnet:2011} investigated the Orowan strengthening at low temperature using the mobility laws described above, first in the case of a square arrays of impenetrable precipitates. Since mobility of edge dislocations are independent of length the additional stress to the lattice friction induced by the periodic array remain controlled by line tension in agreement with the so-called Bacon Kocks Scattergood (BKS) model ~\cite{Bacon:1973}. This quasistatic behaviour contrasts with the case of screw dislocations that exhibit a strain rate dependence. The Orowan strengthening still depends upon spherical size D and spacing L of the precipitates but now the extra stress is significantly decreased with temperature when compared to the BKS model. Due to the length dependence in their mobility, screw dislocations remain mostly straight when forced against precipitates and non-screw arc is formed closed to precipitates. The authors identified two different configurations depending on the relative ratio or the border-to-border spacing $L-D$ and twice the local curvature adopted $2\kappa$. i) For large precipitate spacings such that $L-D > 2\kappa$ , the Orowan stress corresponds to the change of stress induced by the change of screw length. This first interaction is consistent to earlier model  proposed in ~\cite{Louchet:1979,Kubin:2013} from TEM observations. The arc formed by the non-screw segments is not controlled by line tension ~cite\cite{Kubin:2013} but by the production of DK. Monnet and coworkers ~\cite{Monnet:2011}  provided an analytical model assuming that screw and non screw velocities are identical. For small particle spacing $L-D <  2\kappa$, the initially screw dislocation bows out so much that the screw length is considered to be close to the critical DK length, and line tension effects need to be included. In the same study, Monnet and collaborators investigated the low temperature strengthening induced by random distributions of precipitates. To speed up simulations, the initial microstructures were essentially composed of screw dislocations. In steady state conditions, the area swept by dislocation equals $A = \lambda^2 = \bar{L}(\kappa+D/2)$ where $ \bar{L}$ is average particle spacing in the length direction. In these massive calculations, the precipitate microstructure was such that $\bar{L}-D > 2\bar{\kappa}$. The additional stress induced by Orowan particle was therefore derived from the periodic raw model i) as:
\begin{eqnarray}
\Delta \tau_{Orowan} = \tau_P {2k_BT \over \Delta G_0} \ln \left( {l_0 \over \bar{l}} \right)\left(2+ {2k_BT \over \Delta G_0} \ln \left( v^2 \over H^2l_0  \bar{l}  \right) \right)
\end{eqnarray}
with $ \tau_0$ and $\Delta G_0$ the parameters first appearing in the screw mobility in eq. [1.1].  $\bar{l} = 1 / \left( D \rho \left( {\mu b \over \tau_{app} } +{D \over 2} \right) \right)$.

Naamane ~\cite{Naamane:2008} investigated the evolution of forest strengthening at low temperature making use of model latent hardening simulations. To mimic low temperature microstructures, long mobile screw dislocations are interacting with forest dislocations made of screws and may form symmetrical mixed junction. When a junction is formed one of the ternary node is zipping the reaction in the direction of screw motion while the other is zipping in the opposite direction. This may help the motion of one of the mobile arm of the junction, and ultimately leading to an asymmetry in the unzipping process depending on which of the two arm is the more mobile (longest). Interestingly, the average junction length was found to be constant for all temperatures and dislocation densities investigated. The additional strengthening due to forest interactions $\Delta \tau_F$ to the lattice friction was found to decrease as the temperature decreases. Finally, the additional $\Delta \tau_F$ at low temperature was found in qualitative agreement with the model proposed by Louchet et al. ~\cite{Louchet:1979}. Monnet and coworkers ~\cite{Monnet:2013} proposed a bit latter a unified crystal plasticity formalism based on these DDD results that encompasses both the low and high temperature behaviours of bcc metals ~\cite{Monnet:2013}. This model captures most features of the plastic deformation of bcc single-crystals. More details on the peculiar dislocation physics observed at low temperature can be found in ~\cite{Kubin:2013,Monnet:2013}. 
 
In real materials, several strengthening mechanisms are often operating simultaneously such as several reactions pinning dislocations. The question of how to combine the various strengthening associated to define the overall flow stress $\tau_{tot}$ from the separated mechanisms $\tau_c^m$ becomes important and exhibits generally the following form $(\tau_{tot})^k = \Sigma_m (\tau_c^m)^k$ ~\cite{Brown:1972}. The relative strength and density of the dislocation obstacles may dictate the choice of exponent $k$. The quadratic law of mixture where $k = 2$ is only exact when the mechanisms have identical strength but  densities are different, while $k = 3/2$ is exact when the two have identical densities but  strengths are different. Interrestingly, these two extreme cases are rarely very far from each other for a wide range of relative density and strength. Law of mixtures have been identified from DDD ~\cite{Monnet:2006, Queyreau:2010}.
in ~\cite{Monnet:2006}, Monnet has determined the law of mixture associated to the strengthening associated to two different populations of incoherent precipitates in Zr Alloy system. Only a mixture law close to the one proposed by Brown was capable of predicting the set of  DD results:
\begin{eqnarray}
\tau_{total} = \left({ \rho_1 \over  \rho_1+ \rho_2 } \right)\tau_1 + \left({ \rho_2 \over  \rho_1+ \rho_2 } \right)\tau_2, 
\end{eqnarray}
where $\rho_i$ are area densities of the two families of obstacles. In ~\cite{Queyreau:2010}, law of mixture have been compared and a quadratic law provided the best agreement with various strengthening ranging from forest interaction to spherical impenetrable precipitates. Linear superposition is often considered to be justified for the combination of relatively large or strong obstacle among a population of weak obstacle, say as in the case of solute atoms with forest of precipitates. 

\subsection{Towards polycrystalline plasticity}

Including the formation and presence of grain boundaries (or possibly interfaces) is certainly the new frontier for DDD simulations. This is associated to a number of challenges. i) The presence of internal interfaces makes the resolution of the stress and displacement fields much more difficult in comparison to the case of the single crystals, as exposed in section 1.2.4.  ii) In addition to the remote interaction existing between the stress fields of crystals dislocations and interfaces, a set of contact interactions are possible ranging from impeding, partial absorption or recombination with the interface defects and direct or indirect reemission of dislocations on the other side of the interface. iii) The plastic behaviour may then differ from the one observed in single crystal with dislocation accumulation at grain boundaries, a different strain localisation, stress concentrations, and modified hardening. However, since the overall behaviour iii) depends on the stress state i) and elementary mechanisms that are operating ii), the full picture is certainly still out of reach. 

Remote interactions between the GB and the dislocation can lead to the repulsion or attraction. In addition to their blocking by GB, dislocations may be partially absorbed by dislocations and recombined with GB dislocation. Burgers vectors of GB dislocations may belong to DSC lattice ~\cite{Sutton:1995} which is defined from the dichromatic structure of the GB. After absorption, a new dislocation can eventually be emitted in the adjoining crystal, leaving behind a residual dislocations. Since this evolves complex core rearrangements with closely spaced GB dislocations, the mechanical picture is rather difficult to assess. In \cite{Koning:2003}, a qualitative line tension model was proposed to evaluate the chance of a given slip system to be transmitted across several simple CSL GBs. However, as direct transmission involves complex rearrangements and large stresses, indirect transmission through local stress concentration activating another slip system in the adjoining grain could be more efficient. For polycrystals with micrometer grain size, addition mechanisms such as GB source nucleation and GB sliding are suggested. 

In this context, interfaces have been modeled in DDD in different ways such as impenetrable boundaries \cite{Sansal:2009,Aifantis:2009, Zhang:2014, Stricker:2016} mostly impenetrable boundaries with transmission permitted according to a simplified criterion \cite{Fan:2015}, explicit GB structure as simple dislocation arrays in a continuous matrix \cite{Liu:2011, Liu:2012, Devincre:2009, Burbery:2017}, or gamma/gamma prime coherent interfaces using DCM modelling in Ni based super alloys \cite{Vattre:2014, Jamond:2016}. As it was the case for single-crystal simulations, isotropic elasticity is mostly employed. The geometry of grains and surrounding grains can have a drastic impact on the observed behaviour, from bi- or tri-crystals \cite{Aifantis:2009,Zhang:2014,Devincre:2015}, perfectly periodic cubic grains \cite{Fan:2015,Jiang:2019}, to four cubic or rectangular grain aggregates \cite{Jiang:2019}, to 16 dodecahedral grain aggregates \cite{Sansal:2009}.

One of the first impact of the presence of GB, is the reduction of dislocation mean free path. A number of studies have investigated the accumulation of dislocations at GB and the resulting backstresses. In \cite{Daveau:2012, Devincre:2015} a few sets of large and semi-infinite bicrystals have been investigated with DDD, where one or the two grains are plastically deformed in single glide. The selected grain orientations matched the actual orientation found in a tricrystal of Ni. The accumulation of Geometrically Necessary Dislocations (GND) at the GB is not the textbook pile-up picture as the initial microstructures are made of prismatic loops and cross-slip is operating.  The authors proposed a modified pile-up expression reproducing DDD results including the GND density decrease with normal distance and slip plane orientation. The total GND density is then included in modified CP model following \cite{Kocks:2003} i) this density is removed from the Taylor expression, ii) An additional storage is expressed as $d\rho_{GND}^i/d\gamma^i = 0.5 / br$, with $r$ the distance from the GB, iii)  The GND presence is also at the origin of a backstress that is included in the yield criterion. In \cite{Aifantis:2009, Zhang:2014}, DDD simulations are employed for the deformation of a tricrystals in Al. Two impenetrable and pure twist  [001] GB separate the series of three crystals. Deformation is imposed along [001] consistent with multislip condition. The initial microstructure is made of FR sources and different lengths were tested. The distribution of GND at GB is analysed in terms of dislocation density tensor calculated over slices of the simulation box. Alternatively to a CP model, DDD results are used in this study to implement a Gradient Plasticity (GP) model (see corresponding chapter in this book). Few parameters are fitted from the confrontation of  DDD simulations with the 1D resolution of the GP. The backstress induced by GND dislocations is a natural outcome of the model. Interestingly, the internal length associated to the higher order term is not constant but scales with the controlling length-scale e.g. the FR source length or the forest spacing. Besides, while GB are impenetrable in the DDD simulations, a better agreement is found when the GB are modelled as deformable interface in the GP with an efficient GB strength defined from dislocation activity within the grains (see also \cite{Stricker:2016}). The same DDD simulation have been also compared to a different GP model in \cite{Bayerschen:2015}. Both CP and GP models provide a self consistent in the sense that they match rather well the DDD simulations. It would be interesting to test their predictability to other cases. 

The transmission of crystal dislocations through a LAGB  is investigated in \cite{Liu:2011, Liu:2012} in bcc Fe. in \cite{Liu:2011}, a $<111>$ pure tilt symmetric LAGB is modelled by edge dislocations separated by 100$b$. Several infinite mobile segments are considered to test the six binary contact reactions existing among $1/2<111>\{110\}$ slips systems (see section 1.2.1). Positive and negative incoming dislocations are considered and the GB is not loaded. The outcome of these simulations are rather different from the latent hardening simulations as the dislocation line directions are imposed by the boundary plane. As a results, no weak reaction i.e. edge and asymmetrical mixed junction are formed leading to an easy transmission of the mobile segment, and the strongest interaction is found, as expected, with the negative collinear mobile segment and a symmetrical edge junction formation. For the latter case, and depending on the applied stress a recombination is observed into an hexagonal dislocation network consistent with a mixed GB. Following atomistic indications \cite{Terentyev:2010} of mobility of binary $<100>$ junctions is included and could potentially impact stability of ternary junctions. A follow up study \cite{Liu:2012} investigated the interaction between mobile segments with mixed tilt-twist LAGB still in bcc Fe. The LAGB is modelled as 2 sets of $1/2<111>\{110\}$ dislocation relaxing into a hexagonal array and forming a $[0 1 0](1 0 1)$ binary reaction. The interaction with an infinite segment now lead to two possible binary junctions and a ternary reaction, which can be axial or delocalized zig-zags (see section 1.2.1). Different loading conditions of the GB were considered. Various levels of transmission stress were observed as function of the nature of the interaction, and desintegration was sometimes observed before dislocation penetration. It is interesting to note the large stress level associated to this LAGB of respectively $0.57\deg$ and $0.3\deg$ of misorientation, respectivelly. It is expected that GB dislocation spacing decreases with an increase in the misorientation angle, leading to possible higher stress level for dislocation transmission. Also, these model configurations are perfectly periodic ad it would be interresting to see the impact of existing defect on the transmission stresses.

\begin{figure}[h]
\includegraphics[scale=0.5,bb=0 0 100 100]{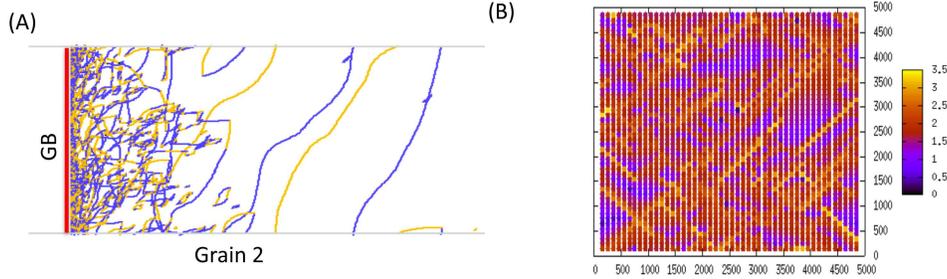}
\caption{Cross view of dislocation accumulation at impenetrable GB in a fcc Al bicrystal after 0.2 plastic strain. The GB is a $\Sigma$5 pure twist. Corresponding mapping of the stresses $\sigma{zz}$ induced by dislocations at the GB with $\mathbf{z}$ the normal to the GB plane. Stresses are in GPa and distance units in $|\mathbf{b}|$. }
\label{cyclic}
\end{figure}

In the context of crack initiation at GB, Shi and Queyreau ~\cite{Shi:2020} investigated the stress induced by crystal dislocation at impenetrable GB in fcc Al. Initial microstructures are made of prismatic loops. The stresses are calculated using the non-singular dislocation theory ~\cite{Cai:2006} on various microstructure and few GB planes. This means that this is not exactly the Cauchy stress and it constitutes an upper bound as elastic relaxation are neglected in DDD (as opposed to DDD-FEM methods). As expected stress concentrations are rapidly going up with the accumulation of dislocation at the boundaries (see figure 1.2). For a fine mapping of the stresses at the GB, the distribution of stress values seem to converge towards a unique log-normal distribution. After few percent of deformation the stresses reach few GPa demonstrating that dislocation are the necessary ling to atomistically informed debonding formalism (the so-called Traction Separation Law) and macroscopic cohesive zone models. 

Finally, the so-called Hall-Petch size effect has been the subject of several investigation. It states that the flow stress of a polycrystals $\sigma$ scales as the grain size $D$ as: 
\begin{eqnarray}
\sigma = \sigma_0 + { k_{HP} \over D^n}
\end{eqnarray}
where $\sigma_0$ is a reference stress, $k_{HP}$ is a poorly known constant, and exponent $n$ is close to 0.5 but may slightly differ from it. Desansal and coworkers \cite{Sansal:2009} investigated the plastic flow onset of 16 grain aggregates of 14 faces polyhedra with DDD. Grains have random orientations with a Taylor factor of 3.05, and are separated by impenetrable grains boundaries. Consistent with early plastic deformation stages, grains are mostly deformed in single slip. The initial microstructure made of sources, lead to pile-ups at the GB but CS of screws rapidly redistribute dislocation activity within grains. A Voce law is used to model plastic behaviour and a grains size dependence with an exponent $n\approx$ [-0.5, -1] is found for  grain sizes in the range 1-2.5 $\mu$m. The authors defined a $k_{HP}$ from DD simulation from an argument based on dislocation storage at GB and that was scaling with total deformation. Fan and coworkers investigated size effects in periodic grains in hcp Mg \cite{Fan:2015}. The c-axis was rotated  with respect to the z-axis of the simulation to simulate a strong basal texture component, and the crystal was deformed along various multislip directions with a large strain rate. the initial microstructure is made of FR sources on basal <a>, prismatic, pyramidal I or pyramidal II <c+a> Burgers vector, with different Peierls stress on each systems. GB are impenetrable with a simplified transmission criterion. A size effect was observed for the different loading direction and activated systems, with $n$ in the range [-0.4, -1.16] and $k_{HP}$ depending on the nature of the activated slip. Recently, Jiang and coworkers investigated the impact of the grain shape and grain environment on the HP effect in fcc Cu \cite{Jiang:2019}. Initial microstructures are made of prismatic loops in impenetrable grains. Simulations encompass varied configurations, with periodic rectangular grains with varied orientations including a single glide [135], and two multislip [001] and [110], four grain aggregates with two random orientations and a [001], and [110] orientations, and different grain shapes with different aspect ratios mimicking cube, lath and needle geometries. The reference stress $\sigma_0$ is defined the yield stress on an ideal Taylor polycrystal. At plastic onset, the grain size dependence is found with an exponent of $-0.5$ with grain sizes $D$ from 1.25 to 10 $\mu$m. For equiaxed periodic grains, $k_{HP}$ is mostly independent of the number of slip systems with $\approx 4 /bD $. $k_{HP}$ depends strongly on the grains shape, and a shape factor is introduced by the authors to relate to the equiaxed grains, finally the the behaviour of a four grain aggregate is consistent with the average behaviour of the constituent.

\subsection{Cyclic deformations}

\begin{figure}[h]
\includegraphics[scale=0.35,bb=0 0 100 100]{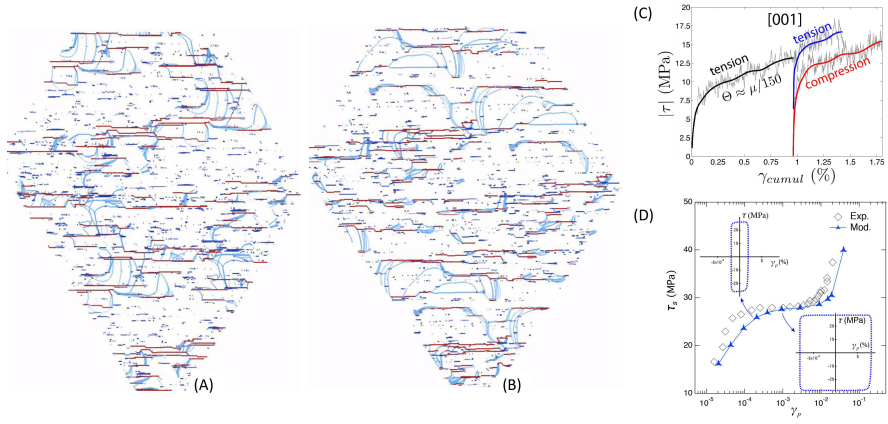}
\caption{Strong tension-compression asymmetry in the plastic deformation of pure fcc metals. A) and B) are sequences of dislocation motion during a Bauschinger experiment using DDD. A Ni single crystal is deformed along [112] direction. Primary dislocations in light blue form junctions in red with the second active slip  system. Thinfoils of 0.2 micrometer thick are extracted from the same region, slightly before and after strain reversal. Note the changes in the paths taken by dislocations and the less continuous propagation of dislocation during compressive deformation. C) Typical Bauschinger curves obtained from DDD in the case of [001] deformation. D) Comparison of the predictions obtained from DDD and the modified CP model with the saturation stress observed experimentally in Cu single crystals ~\cite{Mughraby:1978} (see text for more details).}
\label{cyclic}
\end{figure}

Previous sections were concerned with monotonous loading, however plastic deformation is known to be strongly dependent upon deformation history. One way to assess this aspect of plastic deformation is through cyclic deformations and the corresponding hysteresis curves, starting from the simplest Bauschinger experiments. Initially isotropic materials, exhibit a rounded onset of plastic flow over a transient in compression and sometimes a permanent recovery, when a first loading in tension was applied. This hysteresis behaviour can be found in a multitude of crystalline materials ranging from single-crystals, to polycrystals and in pure or alloyed materials ~\cite{Mughrabi:1978, Mughrabi:2010, Li:2011}. While the ubiquity of this behaviour is clear, its origins are much less evident. Single-crystals are an elementary part of  polycrystals, the cyclic deformation in these systems is even more confusing. Many of the current explanations reside in the easy dissolution of existing dislocation microstructures formed during the initial loading, or in the existence of polarised long range internal stresses. However, there was no clear explanation of the former in terms of elementary mechanisms until recently. The latter explanation is typically associated to the so-called composite model ~\cite{Mughrabi:1988}, according to which the microstructure is made of hard dislocations walls, mostly composed of edges, and soft interior cells where most of the plastic flow occurs. Such a polarised microstructure, lead to longe range internal stresses LRIS, which shall oppose or help the forward or backward motion of dislocations. However, recent X-Ray microdiffraction question the very existence of polarised LRIS in single crystals ~\cite{Kassner:2009, Kassner:2013} . 

When now focussing onto cyclic deformation of single crystals, hysteresis curves evolve with the number of cycles in a complex manner as function of loading direction, the material and and the amount of applied plastic strain per cycle (see review ~\cite{Li:2011}). The maximum stress increases each cycle until reaching a saturation value $|\tau_{sat}|$. In single glide condition, there is a range of intermediate strain per cycle $\gamma_p^{cy}$, for which the saturation stress is almost constant (or slightly increasing), for larger strain per cycle just below $1\%$, double slip eventually activated and saturation stresses increased with the $\gamma_p^{cy}$, while for multislip conditions, the saturation stress increases continuously over the range of typical $\gamma_p^{cy}$. In this context, tridimensional DDD simulations are certainly very attractive as they can provide elementary mechanisms and average the collective  behaviour of dislocations during cyclic deformation. However, despite the constant progress in computing performance and numerical algorithms, only few dozens of cycles are typically accessible to DDD. Therefore, a number of strategies have been proposed to connect to latter stages of deformation by either extrapolating the dislocation patterns or through multiscale approaches. 

In ~\cite{Depres:2008}, Déprés and coworkers considered the cyclic response of a single  dodecahedric grain in single slip conditions in AISI316L stainless steel, to 20 cycles at 1.5 $10^{-3}$ strain. The initial microstructure is made of sources, with cross-slip enabled and the grain boundaries are impenetrable to dislocations, and will leave the place to the presence of many dipolar configurations after the 20 cycles. A distribution of dipole height is then obtained from the simulations and  is assumed to represent the effective opposing of promoting effect to subsequent dislocations. The authors then modified a crystal plasticity model where the dipolar distribution appears in a backstress term. After fitting of a couple of additional parameters, the CP calculations qualitatively captures the evolution of hysteresis curves with cumulated strain for one strain increment in the case of single glide conditions. At the heart of BE and cyclic deformation, lies the question of how much of plastic deformation is reversible by an opposite deformation. This was discussed by Queyreau and Devincre ~\cite{Queyreau:2009} in the case of very model simulation where an infinite dislocation interact with random particle of various strength, where the mean free path of dislocation is directly related to the number of precipitates being swept by dislocations. 

More than in other applications, reproduction of the peculiar patterns observed during deformation have been the motivation to a number of DDD investigations, in an effort to directly relate microstructures to observed deformation behaviour, effort that remains mostly unsuccessful. In single glide conditions, the saturation stress in the plateau is rather close to the onset of dynamic recovery $\tau_{III}$ in monotonic conditions ~\cite{Mughrabi:1979}, where motion and interactions among screw dislocations confined in cells seems evident key. The existence of the plateau, coincides also with the apparition and development of ladder structures also called Persistent slip band. However, it turned out that the saturation stress in multislip conditions is rather far from $\tau_{III}$, despite the existence of marked patterns ~\cite{Li:2011}. Therefore, the direct link between dislocation patterns and saturation remain unsure. In ~\cite{Shin:2005, El-Awady:2008, Hussein:2016}, few cycles of deformation under single-glide conditions are performed to qualitatively assess the development of patterns. Other simulations have model the microstructure by long edge dipolar walls between which screws navigates and interacts. PSB can be made of long infinite and rigid edge dislocations ~\cite{El-Awady:2008} to more realistic and flexible dipolar loops ~\cite{Erel:2017}. Another motivation to simulate the building up of dislocation microstructures is to investigate the specific surface roughthening associated to cyclic experiments. Dislocation activity localization associated with PSD formation is correlated with dislocation exiting though out crystal surface in the form of band extrusions, or accumulating a grain boundaries, possibly leading to crack initiation. Slip band on surface has been investigated in ~\cite{Depres:2006, Hussein:2016}.

Queyreau and Devincre ~\cite{Queyreau:2020} revisited recently the Bauschinger Effect (BE) and cyclic deformation of fcc single crystals (see figure 1.3). It turned out that in DDD simulations a relatively large BE could be obtained but internal stress are shown to be absent. Two original mechanisms are identified from the simulations: i) the stability of junctions formed during the first loading is actually asymmetric, and easily destroyed in the beginning of compression deformation, ii) the avalanches leading to the storage of dislocation segments now occur in region that have been swept during tension, leading to an unwinding of stored segments and loop collapses as in particle strengthened materials ~\cite{Queyreau:2009}. The interaction matrix and MFP of dislocations are weakened by means of two sigmoidal functions $r_a$ and $r_{MFP}$ that state the competition between promoting effects (scaling with $\Delta \rho_p$ stored during forward motion) and finite amount of reversible strain (scaling with $\gamma$ in backward motion). These functions are obtained from a comprehensive set of DDD simulations. These results are extended to larger scale through a modified crystal plasticity model that is solely based on the DDD parameters or known monotonic parameters. Other required parameters are provided by former DDD simulations or known experimental results. This continuous model was compared to most available data existing on fcc single crystals and was successful in predicting from the DDD results most of the features of the cyclic deformations. Saturation of the flow stress with the increase of cycle number occurs naturally. The saturation stress is known to depend upon the amount of plastic strain per cycle and the loading direction and materials. It is interesting that to point that here, there is no requirement to back-stresses or need for the existence of a pattern. The controlling effects are the statistical behaviour of short range reactions, and to the first order, patterning do not seem to play major role in the behaviour, as this is the case for monotonic deformations \cite{Devincre:2008}. Ultimately comparison with data obtained on polycrystals with micrometer grains clearly shows that these effects are controlling cyclic behaviour.

\section{Conclusion and future work}

At the end of the present review, a number of remarks become obvious. First, the growing number of applications clearly demonstrates that DDD has become a very versatile tool. Second, DDD can provide both qualitative and quantitative results, but quantitative data are unsurprisingly more difficult to obtain and often require other inputs or coarse graining at other physical scales, whether simulations or experiments. Third, an important part of the papers presented here are actually not DDD investigations, and was required to justify DDD inputs or outcomes. As Materials Science is a mature field, DDD simulations should find a natural place between tradition and innovation.  

Next, we highlight a number of objective that we intend to follow.
\begin{itemize} 
\item{Simulations should continue to benefit from the constant progresses in terms of analytical or semi-analytical models (e.g force evaluations) or more efficient numerical algorithms. The efficiency gain associated to these progresses should then be objectively shown and compared to previous solutions at fixed numerical errors.}
\item{Thanks to the constant improvement of algorithm efficiency and increase in computing power, more realistic configurations can be considered, in particular, including elastic anisotropy may lead to significant differences in the collective behaviour of dislocations.}
\item{DDD simulations are certainly complicated scientific codes and the subject of constant developments. A growing need within the DDD community is certainly the definition of a set of well-posed benchmarks that could shared among the community. Indicators of the convergence of force of dynamics evaluations could also be defined.}
\item{Large scale DDD simulations provide a natural sampling of the collective behaviour of dislocations. There is however still a lot of effort to provide on the statistical analysis of dislocation microstructures, dislocation kinetics and the link between the two, if any. This is particularly important when coarse graining DDD data to larger scales. }
\item{With a number os success regarding the understanding of single crystal behaviour, the new frontier to DDD simulations is certainly the polycrystal system. Due to the complexity of the GB configuration and structures, a close link with atomistic simulations will certainly be more than ever necessary.}
\end{itemize}

\bibliographystyle{ouvrage-hermes}
\bibliography{11.chapter1}


\backmatter

 \nocite{*}
\printindex

\end{document}